# Multifunctional Metamirrors Involving Simultaneous Linear-polarization Conversion and Focusing

Fei Ding*, Yiting Chen, Yuanqing Yang, and Sergey I. Bozhevolnyi

SDU Nano Optics, University of Southern Denmark, Campusvej 55, DK-5230 Odense, Denmark

These authors contributed equally: Fei Ding, Yiting Chen. Correspondence and requests for materials should be addressed to F.D. (email: feid@mci.sdu.dk)

Reducing a set of diverse bulk-optic-based optical components to a single ultrathin and compact element that enables the same complex functionality has become an emerging research area, propelling further integration and miniaturization in photonics. In this work, we establish a versatile metasurface platform based on gap-surface plasmon meta-atoms enabling efficient linear-polarization conversion along with the complete phase control over reflected fields. Capitalizing on the meta-atom design, multifunctional metamirrors involving linear-polarization conversion and focusing are experimentally demonstrated to generate various kinds of focused beams with distinct phase distributions and wavefronts (including orthogonally-polarized vortex-beam and vector-beam), reproducing thereby the combined functionalities of conventional half-wave plates, lenses, and even spatial light modulators. The proof-of-concept fabricated metamirrors exhibit excellent capability of linear-polarization conversion (> 75% on average) and focusing (absolute average efficiency > 22%) within the wavelength range of 800–950 nm under linearly-polarized excitation. The multifunctional metamirrors design developed in this study opens new avenues in the advanced research and applications targeting photonics integration of diversified functionalities.

# Introduction

Optical metasurfaces, planar artificial materials with subwavelength-scale thickness and unprecedented control over optical fields, have attracted progressively increasing attention and provided promising solutions for cost-effective and high-performance optics in recent years[1-5]. Unlike conventional refractive optics that mold the flow of light through gradually accumulated phase variations during light propagation, metasurfaces can directly modify the boundary conditions for impinging optical waves by arranging meta-atoms with tailored optical response in a periodic or aperiodic manner. In this regard, metasurface-based optical elements have the advantages of lightweight, compactness, planar profiles and relative ease of fabrication, while meeting requirements for delivering increased performance and functionality, therefore offering great potential in the integration and miniaturization of conventional bulky optical components. The recent development of metasurfaces has led to fascinating novel physics and numerous promising ultra-compact devices, such as beam-steering[6-12], surface waves or waveguide couplers[13-19], focusing lenses[20-30], optical holograms[31-35], and polarimeters[36-40].

In addition to the aforementioned metasurfaces designed for specific functionalities, multifunctional metasurfaces that efficiently integrate multiple diversified functionalities into single compact devices have become an emerging research area[41-49]. For instance, metasurfaces can realize polarization-selective distinct functionalities by utilizing strongly anisotropic meta-atoms, with each polarization state being subjected to a different transformation[44-48]. Despite certain achievements, metasurfaces possessing multiple diversified functionalities still remain largely unexplored, especially in the optical range. It is highly desirable that the functionalities of several distinct optical devices, such as polarizers, wave plates, lenses, and spatial light modulators, can efficiently be combined into a single design.

In this paper, we propose a reflective metasurface platform based on gap-surface plasmon (GSP) meta-atoms, enabling efficient linear-polarization conversion along with the complete phase control over reflected fields. Multifunctional metamirrors based on GSP meta-atoms for simultaneous linear-polarization conversion and focusing are experimentally demonstrated, generating focused beams with distinct phase distributions and wavefronts in the near-infrared range, mimicking thereby the combined functionalities of conventional half-wave plates, lenses and even spatial light modulators. The fabricated metamirrors exhibit excellent capability of linear-polarization conversion (> 75% on average) and focusing (absolute average efficiency > 22%) within the wavelength range of 800–950 nm under linearly-polarized excitation. In addition, this approach is further extended to realize multifocal metamirrors with distinctly engineered wavefronts focused at different locations in the same focal plane.

## Results

### Meta-atoms design

The overarching design principle is based on the use of meta-atoms ensuring efficient linear polarization rotation akin to half-wave plates and taking into account the fact that the phase of the cross-polarized reflected field can be changed by $\pi$ radian simply by rotating the corresponding meta-atom over 90°. Therefore, designing two meta-atoms that produce the cross-polarized reflected fields with the phases $\varphi_{cr1}$ and $\varphi_{cr2}$ being different by $\pi/2$ (elements 1 and 2 in Fig. 1a) is sufficient to construct the four-element supercell by rotating elements 1 and 2 over 90° so as to cover the whole $2\pi$ range (Fig. 1a), because $\varphi_{cr3} = \varphi_{cr1} + \pi$ and $\varphi_{cr4} = \varphi_{cr2} + \pi$. One can also refine the phase discretization with more meta-atoms designed with complex structures, which may, in turn, complicate the fabrication.

The corresponding meta-atoms represent gold (Au) nanoantennas tilted by 45° with respect to the $x$-axis, separated by a silicon dioxide ($SiO_2$) spacer layer from a continuous Au film. This configuration supports the GSP resonance formed by the near-field interaction between two metallic layers[50]. When an $x$-polarized ($y$-polarized) wave is normally impinging on the meta-atoms, induced electric-dipole oscillations along the long- and short-axis of the nanoantennas are excited, resulting in the cross-polarized scattering. The cross-polarized scattering is further enhanced by the constructive interference in the multireflection process within the GSP cavity, consequently leading to highly-efficient linear-polarization conversion[41-43,51,52]. Finally, by tailoring the shape and dimensions ($a$ and $b$) of the topmost Au nanoantenna, we can independently control the reflection phase of the cross-polarized reflected light of each meta-atom, achieving eventually the desired phase relationship $\varphi_{cr2} = \varphi_{cr1} + \pi/2$ at the design wavelength of $\lambda = 850$ nm (Supplementary Note 1 and Fig. 1) and enabling the design of the four-element supercell (Fig. 1a). The simulated cross-polarized reflectivity $R_{cr}$, orthogonal linear-polarization conversion ratio $PCR$, and cross-polarized reflection phase $\varphi_{cr}$ of the four selected meta-atoms as a function of wavelength indicate that each of the four meta-atoms functions as a reflective half-wave plate with high-efficiency within the wavelength range of 800–900 nm (Fig. 1b and 1c). Specifically, at the design wavelength of 850 nm, the linear-polarization conversion efficiency is over 95% with the averaged cross-polarized reflectivity of ~ 85% (Fig. 1b). In addition, the reflection phase gradient remains linear across the investigated spectrum, ensuring that the designed phase profile is maintained (Fig. 1c). It should be noted that the coupling between the adjacent meta-atoms are relatively weak in our design, verified by the fact that the beam-deflector composed of these meta-atoms with linear phase gradient can first convert the polarization state of the incident linearly-polarized light and then anomalously reflect the cross-polarized beam with the diffraction efficiency of more than 82%, which shows only ~ 3.5% reduction in conversion efficiency compared with the averaged cross-polarized reflectivity of the uncoupled individual meta-atoms (Supplementary Fig. 2). This relatively weak coupling is ascribed to a relatively large cell period, ensuring that the near-field coupling between the designed elements (that approximately scales as the sixth power of the element separation) remains at a low level.

The facile ability and freedom provided by the proposed meta-atoms to achieve efficient linear-polarization conversion along with the complete phase control over reflected fields allows us to implement multifunctional metamirrors. To validate the excellent performance and versatility of such platform, here we experimentally realized, for the first time, reflective multifunctional metamirrors enabling simultaneous linear-polarization conversion and focusing at the near-infrared wavelength range, which mimics the combined functionalities of bulky half-wave plates, lenses and even spatial light modulators.

**Metamirror for linear-polarization conversion and focusing**

Employing the proposed platform, a broadband metamirror for orthogonal linear-polarization conversion and focusing (PCFM) is first designed and demonstrated. Fig. 2a and Supplementary Fig. 3a,b schematically illustrate the working principle of the PCFM, which converts a normally incident *x*-polarized (*y*-polarized) beam into a *y*-polarized (*x*-polarized) focused beam in reflection. The flat PCFM assumes a hyperbolic phase profile at the design wavelength of $\lambda$ = 850 nm, which is discretized on an *x*-*y* grid and implemented by the meta-atoms (Supplementary Note 3 and Fig. 3c,d). Within the PCFM, each meta-atom is rotated by 45° with respect to the *x*- axis, ensuring that the incident *x*-polarized (*y*-polarized) light is converted to the cross-polarized reflected beam. Fig. 2b shows the scanning electron microscope (SEM) images of the fabricated PCFM sample using standard electron beam lithography (EBL) and a lift-off process (see Methods section for more details), which has a diameter of $D = 50$ μm and a focal length of $f = 60$ μm. Following the fabrication, we characterized the PCFM sample using a custom-built optical setup, which includes a couple of optical components mounted on a three-dimensional (3D) translation stage (see Supplementary Note 4 and Fig. 4 for more details of the measurement setup). By moving all the mounted components together along the axis of light propagation (*z*-direction) gradually, the intensity profiles at different *x*-*y* planes are recorded in sequence.

In Fig. 2c-e, we plotted the measured cross-polarized intensity profiles at different wavelengths when the metamirror is illuminated with an *x*-polarized Gaussian beam at normal incidence. It is clearly observed that the cross-polarized reflected beam is well focused at the corresponding focal plane over a broadband spectrum range (the left panel of Fig. 2c-e and Supplementary Fig. 6). Additionally, the measured intensity distributions along the horizontal (*x*) and vertical (*y*) line cutting through the center of the focal spot are in good agreement with diffraction-limited focal spot profiles, which proves the diffraction-limited focusing performance of the metamirror. By stitching all captured images together, the polarization-converted light intensity profiles on the *x*-*z* plane is obtained and displayed in the right panel of Fig. 2c-e, revealing the revolution of the focusing phenomena for the cross-polarized reflected beam.

Besides the good focusing characteristics, this PCFM features excellent capability of orthogonal linear-polarization conversion. At the design wavelength of $\lambda = 850$ nm, the measured *PCR* is found to be ~ 86%. Remarkably, the good linear-polarization conversion performance is sustained over the wide spectral range of 800–950 nm, indicating the broadband response with good focusing characteristics and highly-efficient orthogonal linear-polarization conversion (Fig. 2f). When the working wavelength moves to a short wavelength that deviates further from the designed value, for instance, $\lambda = 775$ nm, the linear-polarization conversion becomes deteriorated with the *PCR* of only ~ 31%, which can likely be ascribed to the large variations in cross-polarized reflection amplitudes produced by different elements constructing the metamirror. Though the PCFM shows poor ability to convert the linear-polarization at $\lambda = 775$ nm, the focusing quality remains high and the cross-polarized light still assembles a tiny focal spot (Supplementary Fig. 6a).

To further examine the operating performance of designed PCFM, the absolute focusing efficiency, defined as the ratio of the light intensity from the focal spot to the incident intensity, is measured and calculated at the investigated wavelength spectrum, as shown in Fig. 2g. As expected, the polarization-converted beam is dominating in the corresponding focal spots over the wavelength range from 800 to 950 nm, where an averaged focusing efficiency of ~ 22% has been achieved, consistent with the value reported for a near-infrared GSP metasurface focusing mirror[23]. The maximum focusing efficiency (~ 31%) was measured at the wavelength of $\lambda = 875$ nm, which is deviated from the designed value. Even though the measured focusing efficiency is considerably smaller than the averaged cross-polarized reflectivity of the meta-atoms (~ 85%) and the efficiency shows some fluctuations in the operating range, it is substantially larger than the upper limit of ~ 10% estimated for V-antenna metalens with polarization-conversion ability[20,22]. We believe the aforementioned discrepancies are related to the imperfections and the surface roughness of the fabricated nanoantennas, the increased damping related to the Ti adhesion layer between the Au and $SiO_2$ layers, and different excitation conditions in the simulation and experiment. As a final comment, it should be emphasized that the PCFM shows similar performance of linear-polarization conversion and beam focusing when the incident polarization is switched to *y*-polarization, in accord with the symmetry of the meta-atoms within the PCFM (Supplementary Fig. 7).

**Metamirror for linear-polarization conversion and focused vortex-beam generation**

Optical vortex-beam with phase singularity was first discovered in the 1990s, which possesses a helical phase front so that the Poynting vector within the beam is twisted with respect to the principal axis of light propagation[53]. In contrast to the spin angular momentum (SAM) that can take only two values, the orbital angular momentum (OAM) carried by the vortex-beams is unbounded since the topological charge *l* can take an arbitrary value within a continuous range. Thus OAM beams are being considered as potential candidates for encoding information in quantum systems, which can greatly increase the information capacity[54,55]. Currently, metasurfaces have been demonstrated to generate vortex-beams due to their planar configurations and compact footprints[56-62]. Here, we combined the functionalities of three distinct optical devices, a half-wave plate, a lens and a spatial light modulator, into a single design to experimentally realize a flat metamirror for orthogonal linear-polarization conversion and focused vortex-beam generation (PCVFM) under the excitation of linear-polarization (Fig. 3a and Supplementary Fig. 8a,b).

Fig. 3b presents the SEM images of the fabricated PCVFM sample with the topological change of $l = 2$, displaying the 2-fold spiral pattern that is consistent with the 2-fold spiral phase distribution (Supplementary Fig. 8c). Similar to PCFM, all the meta-atoms constituting the PCVFM are tilted 45° with respect to the *x*-axis to guarantee efficient linear-to-linear polarization conversion. The reflected intensity distributions in cross-polarization at different wavelengths were measured using the same setup shown in Supplementary Fig. 4 under *x*-polarized illumination. As shown in the left panels of Fig. 3c-e, the intensity slices along the propagation direction reveal the evolution of the generated vortex-beam, which is slightly focused at the corresponding focal plane of each wavelength. All the intensity patterns on the *x-y* plane have the doughnut-shaped intensity distributions with the intensity minimum at the center, which is the main characteristic of the vortex-beams, that is, intensity singularity. In order to prove the helicity wavefront of the focused vortex-beam, we performed the interference experiment by using a homebuilt Michelson interferometer, where the generated vortex-beam and the tilted Gaussian beam are arranged in the sample and reference arms, respectively (Supplementary Fig. 5). The fork dislocations are clearly observable in the interference patterns (Fig. 3c-e), verifying the phase singularity in a broadband spectrum. If the topological change is increased from 1 to 3, the corresponding doughnut-shaped focal spot expands and the ring radius of the vortex intensity profile becomes larger accordingly (Supplementary Fig. 9). Additionally, the number of the dislocations would be equal to the increased topological changes of the focused vortex-beams.

Since the focused vortex-beam is solely assembled by the converted cross-polarized light, *PCR* can quantitatively gauge the conversion efficiency of the metamirror, which is defined as the ratio of the beam carrying OAM in cross-polarization to the overall reflected power[59]. As depicted in Fig. 3f, the fabricated PCVFM sample exhibit efficient (*PCR* > 80% on average) focused vortex-beam generation in a wide wavelength spectrum ranging from 800 to 950 nm, superior to some reported metasurface-based OAM generators in visible and near-infrared wavelengths[57,59]. Although the *PCR* drops to ~ 35% at the short wavelength of $\lambda$ = 775 nm, the PCVFM sustains the good performance of generating focused vortex-beam with linear-polarization conversion (Supplementary Fig. 10). Subsequently, we measured the absolute focusing efficiency of the total reflected beam and the cross-polarized component (Fig. 3g). Predictably, the vortex-beam is predominantly focused in cross-polarization with the averaged efficiency exceeding 27%, exhibiting at least 10-fold enhancement compared with that in V-shaped antenans[57], while the remaining co-polarized component is strongly suppressed over the wavelength range from 800 to 950 nm. At the design wavelength of $\lambda$ = 850 nm, the measured focusing efficiency of the polarization-converted vortex-beam is found to be equal to ~ 26%.

**Metamirror for vector-beam generation and focusing**

The polarization properties of the demonstrated metamirrors can be designed on demand, beyond the aforementioned homogeneous polarization distribution after orthogonal linear-polarization conversion. In particular, vector-beams, characterized by a spatially varied distribution of polarization vector, have opened new degrees of freedom for numerous unique functionalities and potential applications in nanophotonics[63-65]. Hereafter, we implement the meta-atoms to realize a vector-beam generating and focusing metamirror (VBFM), distinct from conventional schemes requiring lenses and other bulky optical components, such as spiral light modulators[65].

Fig. 4a and Supplementary Fig. 11a,b schematically illustrate the working principle of the VBFM, which simultaneously generates and slightly focuses radially-polarized (RP) and azimuthally-polarized (AP) vector-beams from *x* and *y* linearly polarized beams, respectively. To obtain a VBFM, the selected meta-atoms are not only arranged by undergoing a hyperbolic phase distribution (Supplementary Fig. 11c) but also rotated around their local positions with the angle of $\varphi/2 - \pi/2$, where $\varphi = \arctan(y / x)$ is the azimuthal angle (Supplementary Note 10 and Fig. 11d). As such, the reflected secondary waves from the VBFM will be locally polarized and finally constructively interference at the focal plane to produce focused vector-beams. Additionally, the reflected light can be switched between RP and AP beams by changing the polarization state of the incident light between *x*-polarized and *y*-polarized. Fig. 4b shows the SEM images of the fabricated VBFM sample with a diameter of *D* = 50 μm and a focal length of *f* = 60 μm. Here it should be noted that a high numerical aperture (NA) VBFM capable of tightly focusing vector-beams can be designed accordingly, where a significantly stronger longitudinal field will be formed at the focus for focused RP beam[63].

Subsequently, the VBFM sample was measured using the same setup in Supplementary Fig. 4 and the measured intensity patterns of the reflected beam in the *x-y* plane at the design wavelength of $\lambda$ = 850 nm are displayed in Fig. 4c,d, validating the polarization state of the focused beam. Under *x*-polarized illumination, the focused doughnut profile splits into two side lobes that are aligned along the orientation of the polarizer, manifesting excellent RP beam generation (Fig. 4c). Once the incident light is switched to *y*-polarization, the split lobes remain perpendicular to the polarization axis regardless of the rotation of the polarizer, which indicates that its polarization is oriented azimuthally. The measured intensity profiles of the reflected beams in the *x-z* plane with a linear polarizer placed orthogonal to the incident polarization unambiguously prove the focusing ability of the VBFM and the measured focal length is close to the design value of 60 µm (Fig. 4e,f). As expected from the design, the VBFM demonstrates good performance at other wavelengths (Supplementary Figs. 12-15), and the measured optical bandwidth is around ~ 18% with respect to the design wavelength, superior to the previously reported value of several percent for the dielectric metalenses involving focused vector-beam generation[44,66].

Regarding the VBFM, the conversion efficiency *CE* is critical, defined as the ratio of the pure vector-beam to the reflected power in total. The evaluated *CE* is found to be sufficiently high with the averaging values of ~ 78% and ~ 75% in the wavelength region of 800–950 nm and possesses a drop at $\lambda$ = 775 nm for both the RP and AP beams under *x*- and *y*-polarized excitation, respectively, which is in good agreement with the *PCR* values of PCFM and PCVFM. Since the generated vector-beam is non-ideal, one may notice the slight fluctuations in the intensity distributions of reflected light through the linear analyzer (Fig. 4c,d). Owing to the unwanted co-polarized reflected beam, the intensity distribution of the generated vector-beam is not flat with respect to the different orientations of the linear polarizer in front of the camera (Supplementary Fig. 16). In addition to the hypothesis regarding the imperfect fabrication and uncertainties of the material properties, we believe the non-ideal conversion efficiency is related to the variations in reflectance produced by the different rotations of the corresponding meta-atoms comprising the VBFM, which could result in more fabrication imperfections due to the increased proximity effect. Though the vector-beam generation is not perfect, it is still remarkable that vector-beam is generated and simultaneously focused in a broadband wavelength range of 800-950 nm, with the averaged absolute efficiencies reaching as high as ~ 35% and ~ 34% for the RP and AP beams, respectively. Additionally, the co-polarized reflected beam is greatly suppressed. In particular, we measured ~ 34% and ~ 34 % converted focusing efficiencies for the $x$- and $y$-polarized excitation at the design wavelength of $\lambda = 850$ nm, respectively.

**Discussions**

In this work, we have established a versatile approach to realize efficient linear-polarization conversion along with the complete phase control over reflected fields. By spatially positioning and rotating the meta-atoms, multifunctional metamirrors enabling simultaneous linear-polarization conversion and focusing have been demonstrated to generate various kinds of focused beams with distinct phase distribution and wavefronts, including orthogonally-polarized beam, orthogonally-polarized vortex-beam, and vector-beam, mimicking thereby the combined functionalities of conventional half-wave plates, lenses, and even spatial light modulators. The proof-of-concept fabricated PCFM and PCVFM samples feature excellent capability of orthogonal linear-polarization conversion and focusing within the wavelength range of 800–950 nm with an averaged polarization conversion ratio of ~ 80% and absolute focusing efficiency exceeding 22% under normal illumination with the $x$-polarized beam. Meanwhile, broadband (800–950 nm) focused vector-beam has been experimentally demonstrated with the VBFM sample, where the averaged conversion efficiency *CE* are found to be ~ 78% and ~ 75%, and the averaged absolute efficiencies reach as high as ~ 35% and ~ 34% for the RP and AP beams under $x$- and $y$-excitation, respectively. Owing to the compactness, integration compatibility, and remarkable multifunctional performance, our flat metamirrors approaches pave an innovative avenue to integrate multiple diversified functionalities into a single compact device to efficiently minimize the plasmonic and photonics circuits and systems.

# Methods

**Simulation.** All numerical simulations were performed by using the commercial software Comsol Multiphysics (ver. 5.3) based on finite element method (FEM). For the design of unit meta-atoms (Fig. 1 and Supplementary Fig. 1), we modeled one unit cell by applying periodic boundary conditions on the vertical sides of the cell. The complex reflection coefficients were determined with respect to the meta-atom top surfaces with linearly polarized light being normally incident onto the metasurface. The medium above the metasurface was chosen to be air and truncated using the perfectly matched layer (PML) to minimize reflection. The permittivity of Au is described by the Drude model fitted with the experimental data[67], where the damping constant is increased by a factor of three to take into account the additional losses caused by the surface scattering and grain boundary effects in thin films. The silicon dioxide ($SiO_2$) spacer layer is taken as a lossless dielectric with the constant refractive index $n = 1.45$.

**Sample Fabrication.** All the metamirror samples were fabricated using thin-film deposition and standard EBL techniques. First, a 3-nm-thick Ti, a 130-nm-thick Au, and a 3-nm-thick Ti were deposited onto a silicon substrate using electron-beam evaporation. Then a 110-nm-thick $SiO_2$ space layer was deposited via RF-sputtering. After that, the metasurface was defined using EBL employing a 200-nm-thick PMMA (4% in anisole, Micro Chem) layer at the acceleration voltage of 30 keV. After development in the solution of methyl isobutyl ketone (MIBK) and isopropyl alcohol (IPA) of MIBK: IPA=1:3, a 3-nm-thick Ti adhesion layer and a 77-nm-thick Au layer were deposited subsequently using electron-beam evaporation. The Au patterns were finally formed on top of the $SiO_2$ film after lift-off process.

## Acknowledgements

The authors gratefully acknowledge financial support from the European Research Council, Grant 341054 (PLAQNAP) and the University of Southern Denmark (SDU 2020).

## Author contributions

F.D. and S.B. conceived the ideas for this project. F.D. and Y.Y. performed the numerical simulations. F.D. designed the structure and fabricated the sample. F.D and Y.C. set up the optical system, carried out the optical measurements and analyzed the data. F. D. and S. B. wrote the manuscript with revisions by all. S.B. supervised the research.

## Additional information

Supplementary information is available in the online version of the paper.

## Competing interests

The authors declare no competing interests.

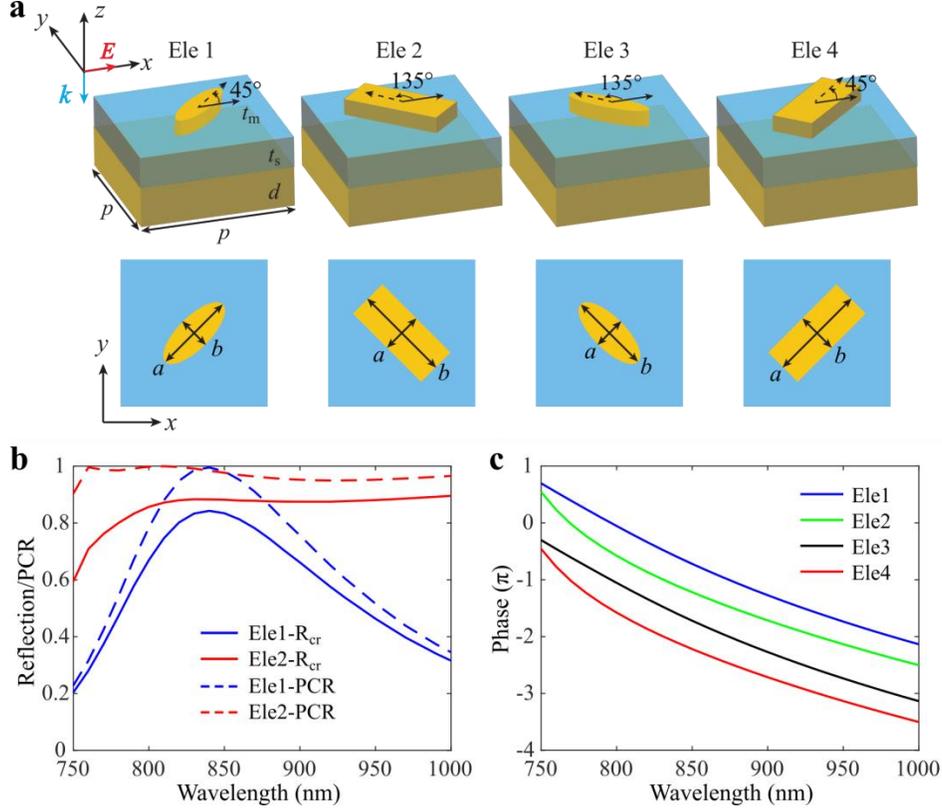

**Fig. 1** Illustration of the meta-atoms constituting the multifunctional metamirrors enabling linear-polarization conversion and focusing. **a** Schematic of the four meta-atoms constructing the metamirrors. The meta-atom consists of an Au nanoantenna on top of a spacer and Au substrate with dimensions $p$ = 550 nm, $d$ = 130 nm, $t_s$ = 110 nm, and $t_m$ = 80 nm. The dimensions of element 1-4 are (1) $a$ = 228 nm, $b$ = 90 nm; (2) $a$ = 155 nm, $b$ = 360 nm; (3) $a$ = 90 nm, $b$ = 228 nm; (4) $a$ = 360 nm, $b$ = 155 nm, indicating that elements 3 and 4 are simply rotated by 90° with respect to elements 1 and 2, respectively. **b** Simulated cross-polarized reflectivity $R_{cr}$ and orthogonal linear-polarization conversion ratio $PCR$ for elements 1 and 2. **c** Simulated cross-polarized reflection phase $\varphi_{cr}$ for elements 1-4 that provide a linear phase variation spanning a $2\pi$ range.

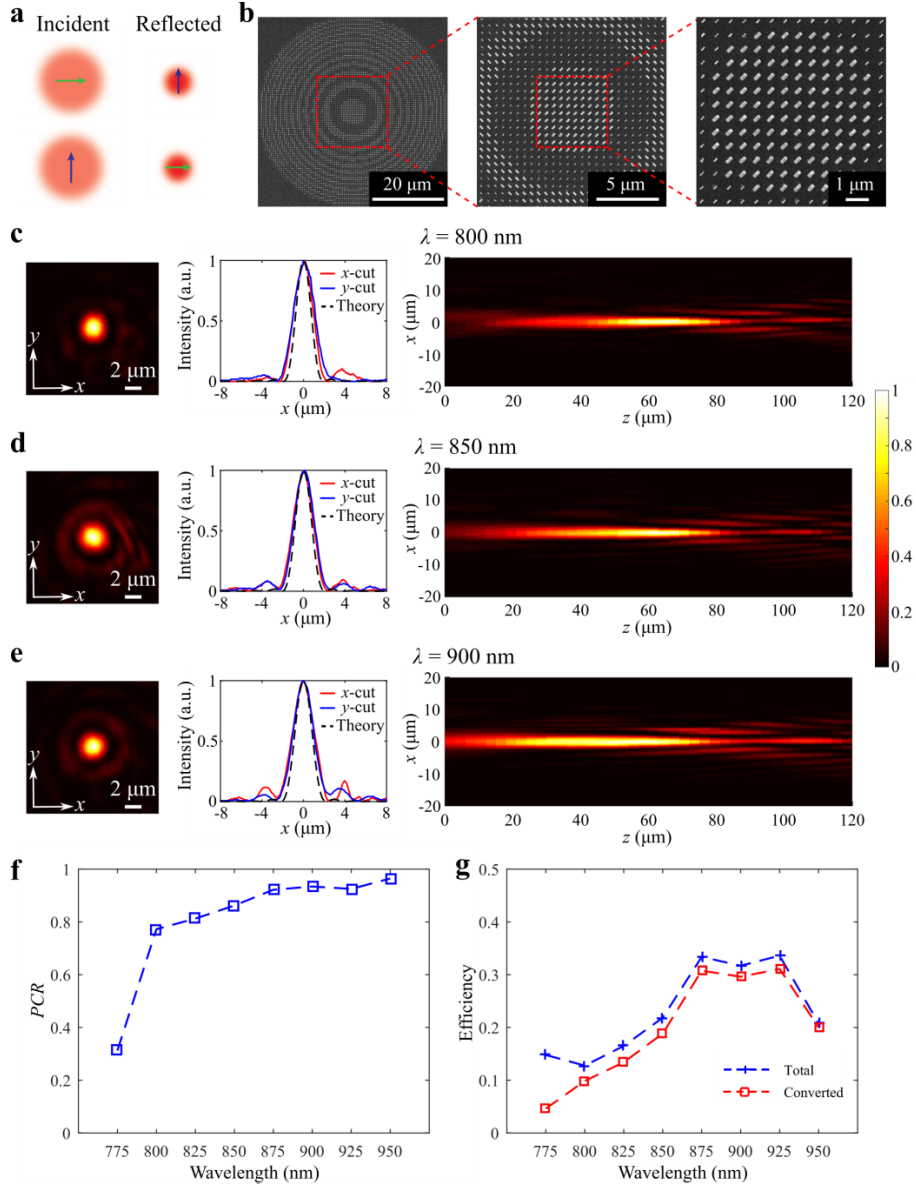

**Fig. 2** Metamirror for linear-polarization conversion and focusing. **a** Schematic illustration of the PCFM, which converts a normally incident *x*-polarized (*y*-polarized) beam to a *y*-polarized (*x*-polarized) focused beam in reflection. Green and blue arrows correspond to *x* and *y* incident linear polarizations, respectively. **b** Top-view SEM images of the fabricated PCFM. **c**-**e** Optical characterization of the PCFM sample at the wavelengths of **c** 800 nm, **d** 850nm, and **e** 900 nm. Left (**c**-**e**): measured focal spot profiles in cross-polarization. Middle (**c**-**e**): measured intensity distributions in cross-polarization along the horizontal (*x*) and vertical (*y*) line cutting through the center of the focal spot in comparison with diffraction-limited focal spot profile. Right (**c**-**e**): measured intensity profiles of the reflected beam in the *x*-*z* plane in cross-polarization. **f** PCR at the corresponding focal plane as a function of wavelength. **g** Measured efficiencies at the corresponding focal plane as a function of wavelength. The total efficiency is defined as the ratio of the light intensity from the corresponding focal spot to the incident intensity while the converted efficiency only considers the reflected beam with orthogonal linear-polarization conversion. The *x*-polarized Gaussian beam is normally incident on the metamirror.

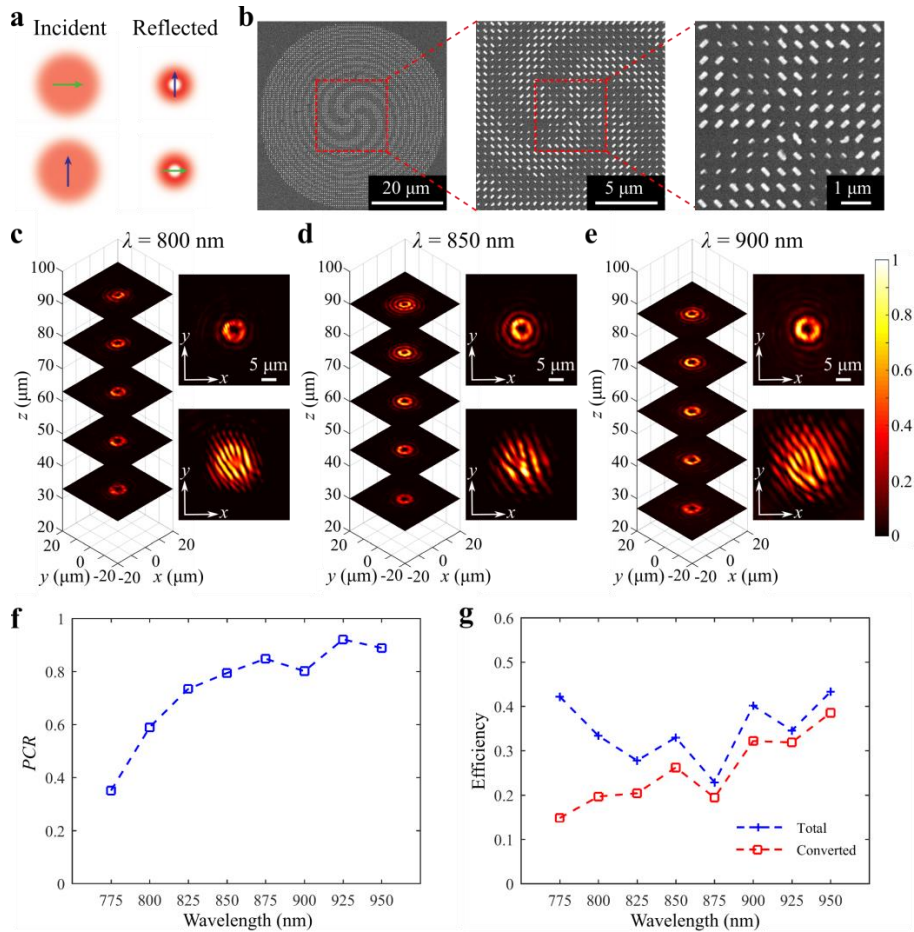

**Fig. 3** Metamirror for linear-polarization conversion and focused vortex-beam generation. **a** Schematic illustration of the PCVFM, which generates focused vortex-beam possessing linear-polarization conversion. Green and blue arrows correspond to *x* and *y* incident polarizations, respectively. **b** Top-view SEM images of the fabricated PCVFM. **c-e** Optical characterization of the PCVFM sample at the wavelengths of **c** 800 nm, **d** 850nm, and **e** 900 nm. Left (**c-e**): focal spots evolution of the generated vortex-beam with topological charge $l = 2$ along the optical axis (*z*-axis) in cross-polarization. Right top (**c-e**): measured focal spot profiles in cross-polarization. Right down (**c-e**): interference pattern of the focused vortex-beam and the copropagating

Gaussian beam when the beam axes are tilted with respect to the other. **f** *PCR* at the corresponding focal plane as a function of wavelength. **g** Measured efficiencies at the corresponding focal plane as a function of wavelength. The total efficiency is defined as the ratio of the light intensity from the corresponding focal spot to the incident intensity while the converted efficiency only considers the reflected beam with orthogonal linear-polarization conversion. The *x*-polarized Gaussian beam is normally incident on the metamirror.

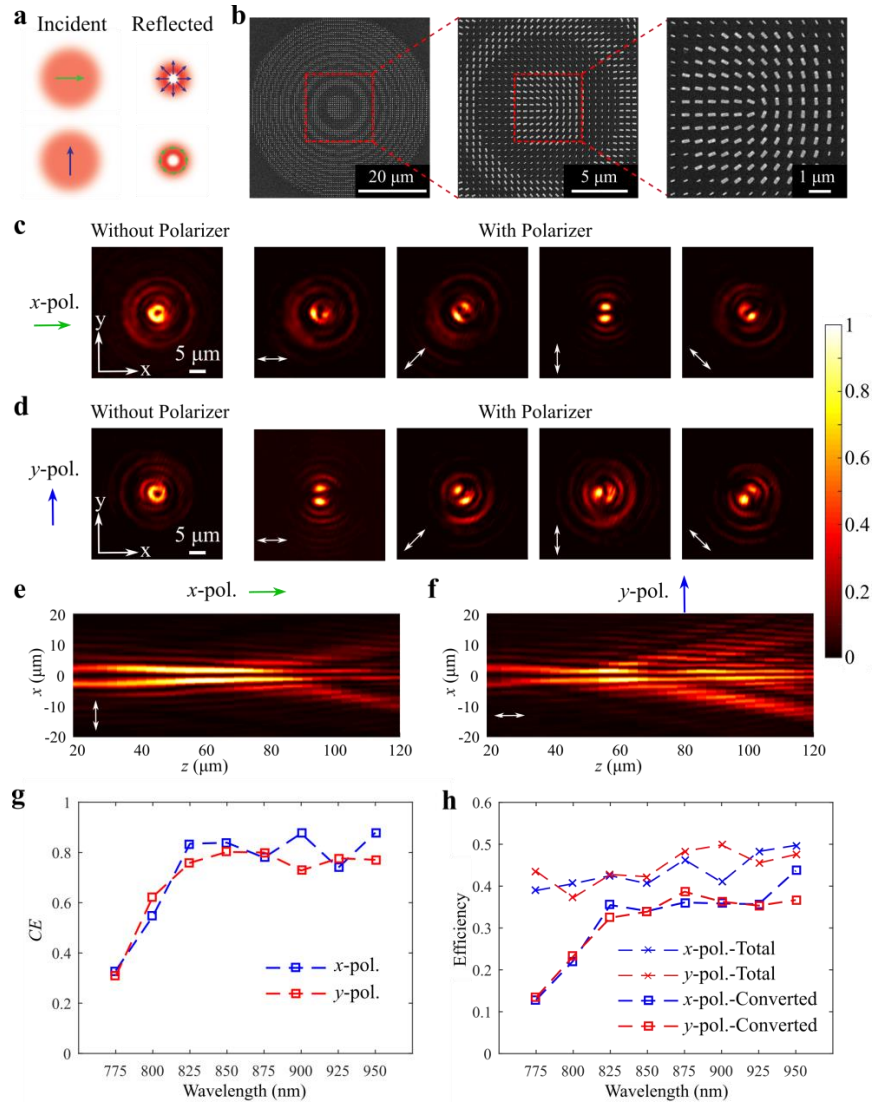

**Fig. 4** Metamirror for vector-beam generation and focusing. **a** Schematic illustration of the VBFM, which simultaneously generates and focuses radially and azimuthally polarized vector-beams from *x* and *y* linearly polarized beams respectively. Green and blue arrows correspond to *x* and *y* incident polarizations, respectively. **b** Top-view SEM images of the fabricated VBFM. **c**,**d**

Measured intensity profiles of the generated vector-beams with and without a linear polarizer placed in front of the camera for **c** *x*- and **d** *y*-polarized incident light, respectively. **e,f** Measured intensity profiles of the reflected beams in the *x-z* plane with a linear polarizer orthogonal to the input polarization for **e** *x*- and **f** *y*-polarized incident light, respectively. **g** Conversion efficiency *CE* at the corresponding focal plane as a function of wavelength for *x*- and *y*-polarized incident beams. **h** Measured absolute efficiencies at the corresponding focal plane as a function of wavelength for *x*- and *y*-polarized incident beams. The total efficiency is defined as the ratio of the light intensity from the corresponding focal spot to the incident intensity while the converted efficiency only considers the reflected vector-beam without the remaining co-polarized part. In **c-f**, the white double-headed arrows show the direction of the transmission axis of the linear polarizer.